\documentclass{PoS}

\newcommand{\be}{\begin{equation}}
\newcommand{\ee}{\end{equation}}
\newcommand{\ba}{\begin{eqnarray}}
\newcommand{\ea}{\end{eqnarray}}

\title{Constraining General Two Higgs Doublet Models by the Evolution of Yukawa Couplings}

\ShortTitle{Constraining General 2HDM by the Evolution of Yukawa Couplings}

\author{Johan Bijnens\\
        Department of Astronomy and Theoretical Physics, Lund University,\\
S\"olvegatan 14A, SE 223-62 Lund, Sweden\\
        E-mail: \email{Johan.Bijnens@thep.lu.se}}

\author{\speaker{Jie Lu}%
         \\
        IFIC, Universitat de Valencia - CSIC, \\
        Apt. Correus 22085, E-46071 Valencia, Spain\\
        E-mail: \email{lu.jie@ific.uv.es}}

\author{Johan Rathsman\\
        Department of Astronomy and Theoretical Physics, Lund University,\\
S\"olvegatan 14A, SE 223-62 Lund, Sweden\\
        E-mail: \email{Johan.Rathsman@thep.lu.se}}

\abstract{We analyse the constraints of the general two Higgs doublet models by evolving the Yukawa coupling constants to high energy under renormalization group.
We consider the appearance of a Landau pole or large off-diagonal Yukawa couplings which cause  tree level flavour changing neutral currents. Our study shows the latter condition can be used to answer how much $Z_2$ symmetry breaking can be allowed in a given 2HDM model.}

\FullConference{Prospects for Charged Higgs Discovery at Colliders\\
                 October 8-11, 2012\\
                 Uppsala University Sweden}

\begin{document}

\section{Introduction}

The two Higgs doublets model (2HDM) is the minimal extension of Standard Model (SM). The introduction of extra Higgs doublet can lead to the tree level flavour-changing-neutral-currents (FCNC), which should be strongly suppressed. Therefore it's necessary to restrict the general two Higgs doublets model in the Yukawa couplings sector. One of common ways is to impose $Z_2$ symmetry for the two Higgs doublets on the Lagrangian \cite{Glashow:1976nt}. Under this symmetry only one Higgs doublet is allowed to interact with each type of fermions so that Yukawa coupling matrix is diagonal at any energy scale.

Recently a more flexible way of avoiding tree level FCNC has been proposed \cite{Pich:2009sp}. This idea simply requires that the two Yukawa couplings
to the Higgs doublets should proportional to each other, so they can be diagonalized simultaneously. This restriction is fine in giving energy scale but the tree level FCNC will be reintroduced in higher energy scale~\cite{Ferreira:2010xe}.
There are also more general proposals which the tree level FCNC Yukawa couplings are suppressed enough, e.g. the Cheng-Sher ansatz \cite{Cheng:1987rs}.

We will study the properties of all these models by taking into account the theoretical and experimental constraints on FCNC. And then we will evolve all the Yukawa couplings to higher energy according to the Renormalization Group Equations (RGE). In some cases the off-diagonal matrix elements which relate to the FCNC grow quickly, which indicates certain assumptions behind the theory are not stable. Those theories are either fine-tuned or incomplete in certain way, e.g. there may be additional particles appearing when going to high energy scale.

\section{The general 2HDM and RGE equations}

One of standard convention to write the two Higgs doublets with the
Goldstone bosons is
\begin{eqnarray}
\Phi_1&=&\frac{1}{\sqrt{2}}\left(\begin{array}{c}
 \sqrt{2}\left(G^+\cos\beta -H^+\sin\beta\right)  \\
 v\cos\beta-h\sin\alpha+H\cos\alpha+\mathrm{i}\left( G^0\cos\beta-A\sin\beta \right)
\end{array}
\right)\,,\\
\Phi_2&=&\frac{1}{\sqrt{2}}\left(\begin{array}{c}
\displaystyle \sqrt{2}\left(G^+\sin\beta +H^+\cos\beta\right)  \\
\displaystyle v\sin\beta+h\cos\alpha+H\sin\alpha+\mathrm{i}\left(
G^0\sin\beta+A\cos\beta \right)
\end{array}
\right)\,.
\end{eqnarray}
Where $G^\pm$ and $G^0$ are the Goldstone bosons to be eaten by the
EW gauge bosons during  EW symmetry breaking, and $H^\pm$ is the charged
Higgs boson. The neutral Higgs scalar can be divided into CP even
scalars $(h, H)$ and CP odd pseudo-scalar $A$.  $\alpha$ and $\beta$  is the
mixing angle between ($h, \ H$) and the two vacuum expectation values (VEV).

For Yukawa coupling analysis, it's convenient to use Higgs basis
\begin{eqnarray}
\label{eq:H1}
H_1& = & \cos\beta\, \Phi_1 + \sin\beta\, e^{-i\theta}\Phi_2\,,\nonumber\\
H_2& = & - \sin\beta\,\Phi_1 + \cos\beta\,e^{-i\theta} \Phi_2\,.
\label{eq:H2}
\end{eqnarray}
Where the nonvanishing VEV is only assigned to $H_1$ which plays the role of Standard Model Higgs doublet while $H_2$ contains the new particles $H^\pm$ and $A$.
The general Yukawa coupling is
\begin{eqnarray}
\label{Yukawa1}
-\mathcal{L}_Y&=&
\overline{Q}_L\widetilde{H}_1\kappa_0^U U_R+\overline{Q}_LH_1\kappa_0^D D_R+\overline{L}_L H_1 \kappa_0^L E_R\nonumber\\
&&+\overline{Q}_L\widetilde{H}_2\rho_0^U U_R+\overline{Q}_L H_2\rho_0^D D_R+\overline{L}_L H_2 \rho_0^L E_R+\mathrm{h.c.}\,.
\end{eqnarray}
Where the subscript "0" stands for flavor basis. In mass basis, the $3\times3$ matrix
$\kappa^F$  is related to diagonal fermion mass matrix by bi-diagonalizing with the unitary matrices $V_L^F,V^F_R$
\begin{eqnarray}
\kappa^F &=& V^F_L\kappa_0^F V^{F\dag}_R = {\sqrt{2}\over v}\mathcal{M}_{ii}^F
\label{eq:MasseigenKR}
\end{eqnarray}
The coupling matrix $\rho^F$ is still general and complex if there are no further restrictions. The Cheng-Sher ansatz suggests
\begin{eqnarray}
\rho^F_{ij} &=& \lambda^F_{ij} \frac{\sqrt{2m_i m_j}}{v}\,.
\label{CSansatz}
\end{eqnarray}
Where the $m_i$ are the different fermion masses, the $\lambda^F$ are expected to be
of $\mathcal{O}(1)$ and should be small enough to suppress FCNC to the
observed level. In EW scale, the $Z_2$ symmetric and aligned models can be treated as special case of Cheng-Sher ansatz.

\section{Numerical analysis}
\subsection{SM input and Constraints}
The most stringent constrains for tree level FCNC is from neutral meson mixing.
The master formula for tree level $F^0-\bar F^0$ mixing can be found in \cite{Atwood:1996vj}.

Using current experimental and theoretical data,  we estimated the bounds for nondiagonal $\lambda^F_{ij}$ with the following assumption of Higgs scalar mass \cite{Bijnens:2011gd}:
\begin{itemize}
  \item $m_h = m_H = m_A = 120$ GeV:
$\{   \lambda_{uc} \lesssim 0.13\,, \lambda_{ds} \lesssim 0.08\,, \lambda_{db} \lesssim 0.03\,, \lambda_{sb} \lesssim 0.05\,\}$;

  \item $m_h = m_H = 120$ GeV,  $m_A = 400$ GeV:
   $\{   \lambda_{uc} \lesssim 0.30\,,   \lambda_{ds} \lesssim 0.20 \,, \lambda_{db} \lesssim 0.08 \,, \lambda_{sb} \lesssim  0.12 \,  \}$.
\end{itemize}
According to these results, we choose $\lambda^F_{ij}\lesssim 0.1$ as a representive value which will be used as generic value later in RGE analysis.

\subsection{$Z_2$ Symmetric Models}

The first example we study is the models with $Z_2$ symmetry, in which the $\tan\beta$ is a physical parameter.
The tree level FCNC is protected by the $Z_2$ symmetry, so the nondiagonal Yukawa couplings stay as zero in any energy scale.
The only thing we can study is to detect the place of Landau pole, where at least one of the Yukawas blow up. Beyond the Landau pole the perturbation theory is not valid anymore.

\begin{figure}
  \begin{minipage}[c]{.25\textwidth}
    \centering
    \caption{The starting value of $\tan\beta$ as a function of the position of the corresponding Landau pole ($\Lambda_{\rm Landau-pole}$) in the different 2HDM types. There are lower limits on $\tan\beta$ for all
    four types (left), but only type II, type III/X and type IV/Y have an upper limit of $\tan\beta$ (right).}
    \label{plot:LPtypes}
  \end{minipage}%
  \begin{minipage}[c]{.45\textwidth}
    \centering
\includegraphics[width=12cm, viewport=50 115 800 460,clip]{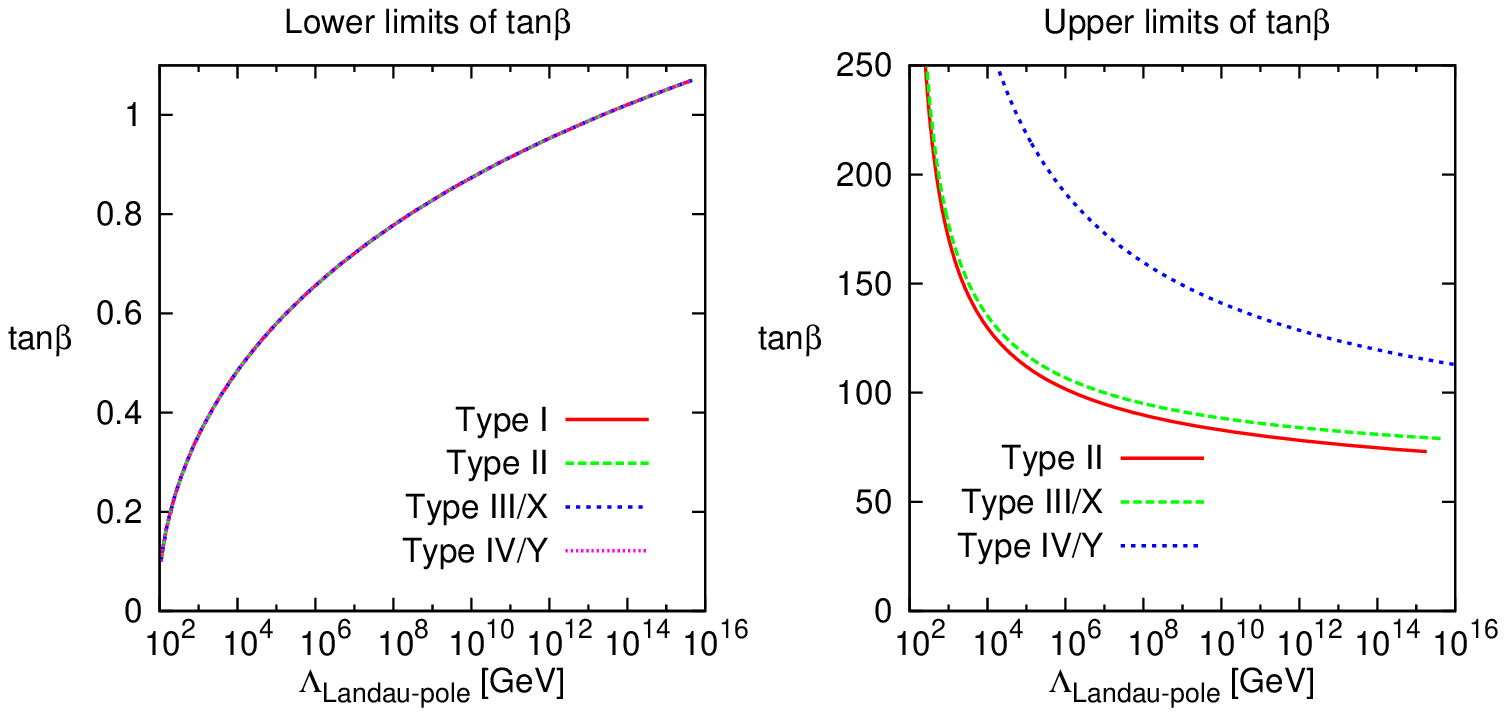}
  \end{minipage}
\end{figure}

In Table \ref{tab:Z2lambda} we show the diagonal $\lambda^F_{ii}$ in terms of $\tan\beta$
for the four different 2HDM types in $Z_2$ symmetric models. The position of the Landau pole depends on the initial value of $\tan\beta$ in EW scale.
In Fig. \ref{plot:LPtypes} we plot the upper and lower limit of $\tan\beta$. The plots can be understood by whether the evolution is driven by $\lambda_{tt}$, $\lambda_{bb}$ ,$\lambda_{\tau\tau}$ or combination of them.
\begin{table}
\centering
\begin{tabular*}{0.4\columnwidth}{@{\extracolsep{\fill}}cccc}
\hline
Type &  $\lambda^U_{ii}$ & $\lambda^D_{ii}$ & $\lambda^L_{ii}$ \\
\hline
I     &  $1/\tan\beta$  & $ 1/\tan\beta$ & $1/\tan\beta$\\
II    &  $1/\tan\beta$  & $ -\tan\beta$        & $-\tan\beta$ \\
III/Y &  $1/\tan\beta$  & $-\tan\beta$         & $1/\tan\beta$ \\
IV/X  &  $1/\tan\beta$  & $ 1/\tan\beta$ & $-\tan\beta$ \\
\hline
\end{tabular*}
\caption{
The diagonal $\lambda^F_{ii}$ in  2HDM models with $Z_2$ symmetry.}
\label{tab:Z2lambda}
\end{table}

\subsection{Z2-breaking Models}
\subsubsection{Aligned models}
In Yukawa alignment model, the two Yukawa couplings $\kappa^F$ and $\rho^F$ are proportional to each other
so they can be diagonalized simultaneously. In this model the $\lambda^F$ is also diagonal in EW scale.
However the nondiagonl element will start to grow when the Yukawas evolve to higher energy via RGE.
Similarly to the Z2-breaking models, we limit the alignment model with three different cases:
\begin{itemize}
  \item Aligned I/II:    $ \lambda^U_{ii} , \quad \lambda^D_{ii}  = \lambda^L_{ii}$
  \vspace{-0.2cm}
  \item Aligned III:    $ \lambda^D_{ii} , \quad \lambda^U_{ii}  = \lambda^L_{ii}$
    \vspace{-0.2cm}
  \item Aligned IV:    $ \lambda^L_{ii} , \quad \lambda^U_{ii}  = \lambda^D_{ii}$
\end{itemize}

\begin{figure}
  \centering
  \begin{minipage}[c]{.45\textwidth}
    \centering
    \caption{The constraints from the Landau pole and non-diagonal $\lambda^F_{i \neq j}$.
  The plot shows the same results for Aligned I/II or Aligned III. The $\lambda^L_{ii}$ only play a very minor role so we don't show the plots for Aligned IV case. The areas inside a given contour are allowed by the requirement of the two condition above.
  The different contours are as follows starting from the center: $10^{15}$, $10^{10}$, $10^{5}$, $10^{3}$,
  and $300$ GeV. The $Z_2$ symmetric case is also plotted as a reference.}
    \label{plot:2DA}
  \end{minipage}%
  \begin{minipage}[c]{.45\textwidth}
    \centering
\includegraphics[width=7cm]{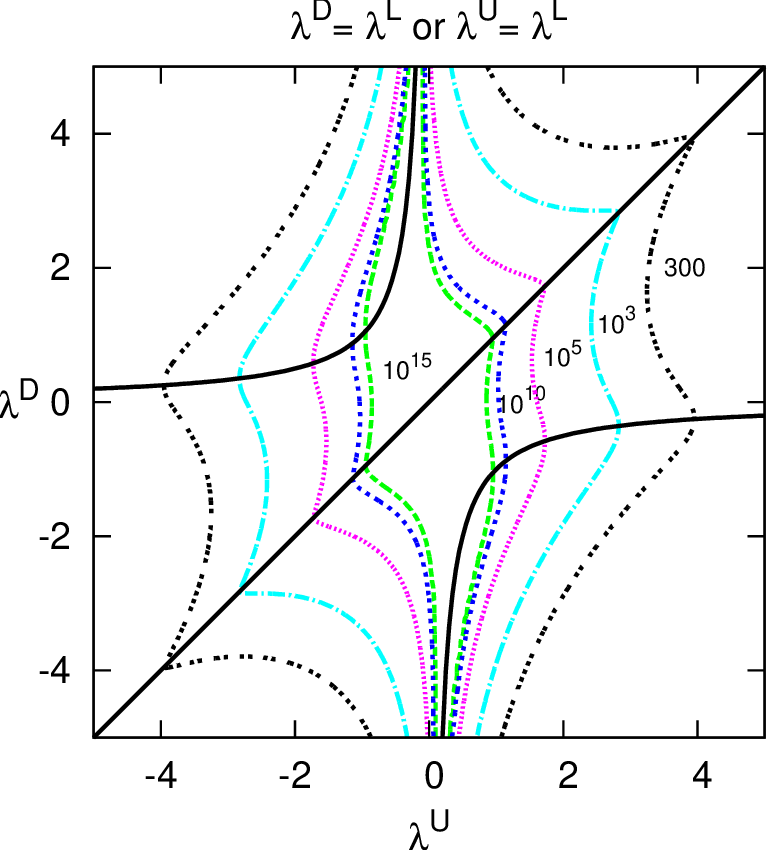}
  \end{minipage}
\end{figure}

In Fig. \ref{plot:2DA} we plot the  energy scale at which the Landau pole or large nondiagonal $ \lambda^F_{i \neq j}  =  0.1$ is encountered as a
  function of pairwise combinations of the starting values for
  $\lambda^U_{ii}$  and  $\lambda^D_{ii}$.
\subsubsection{Diagonal models}
Next we consider the models with $Z_2$ symmetry breaking in either the up or the down sector.
First we break the $Z_2$ symmetry in the up-sector with  $\lambda^D=\lambda_{tt}$ (type I) or
$\lambda^D=-1/\lambda_{tt}$ (type II), and  set $ \lambda_{uu} =\lambda_{cc} $. Then we  break the $Z_2$ symmetry in the down-sector with
$\lambda_{bb}=\lambda^U_{ii}$ (type I) or $\lambda_{bb}=-1/\lambda^U_{ii}$ (type II), and set $ \lambda_{dd} =\lambda_{ss} $.

\begin{figure}
\centering
\includegraphics[width=4.8cm]{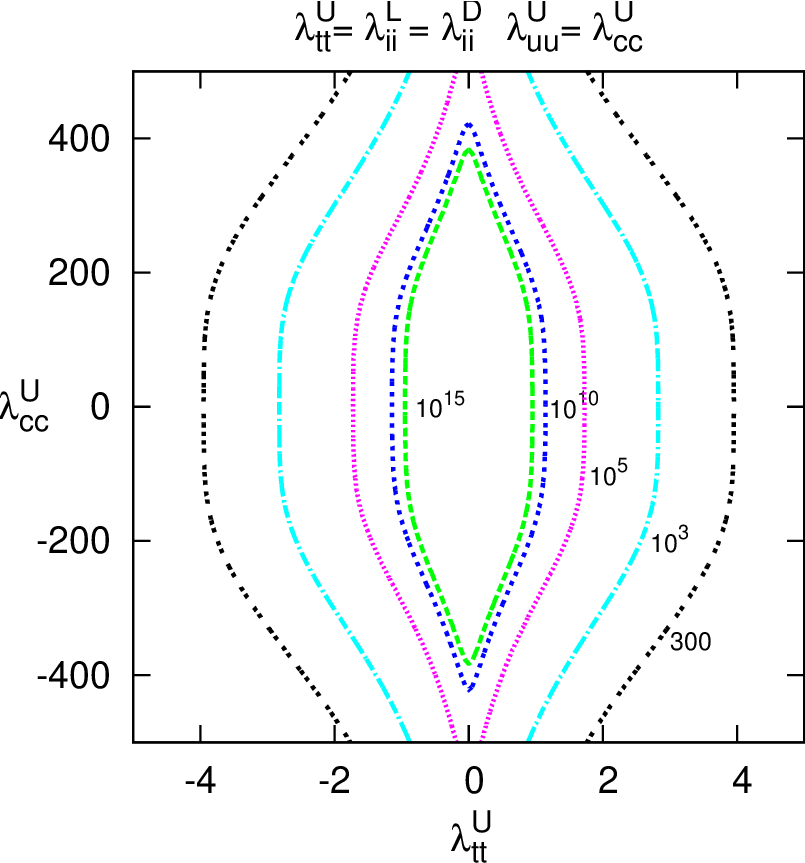}\hspace*{5mm}
\includegraphics[width=4.8cm]{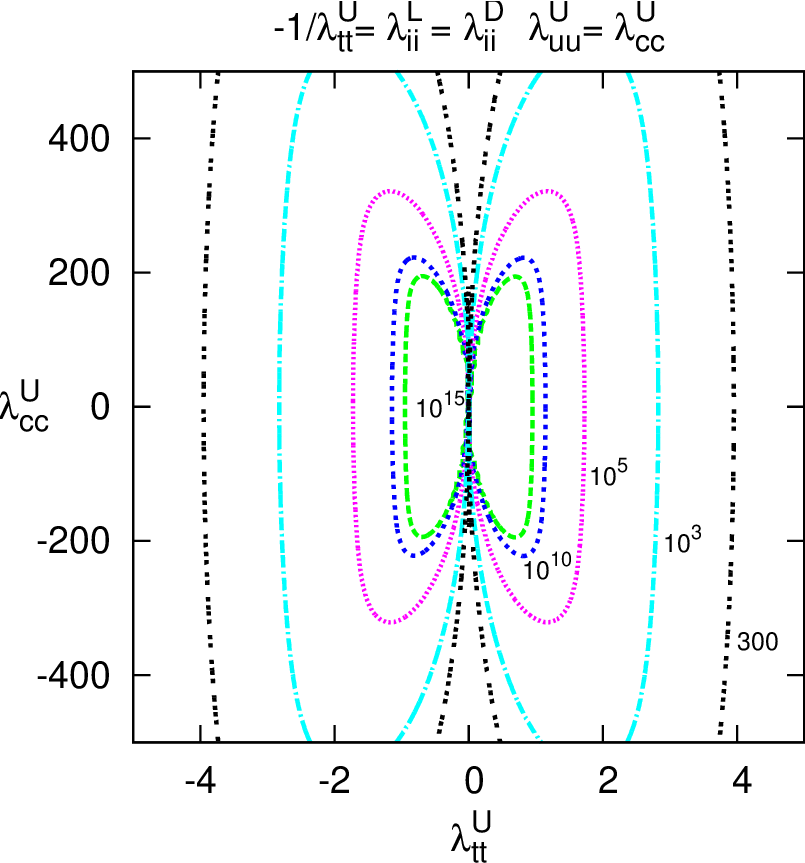}\newline \\
\hspace{-1.5cm}
\includegraphics[width=4.8cm]{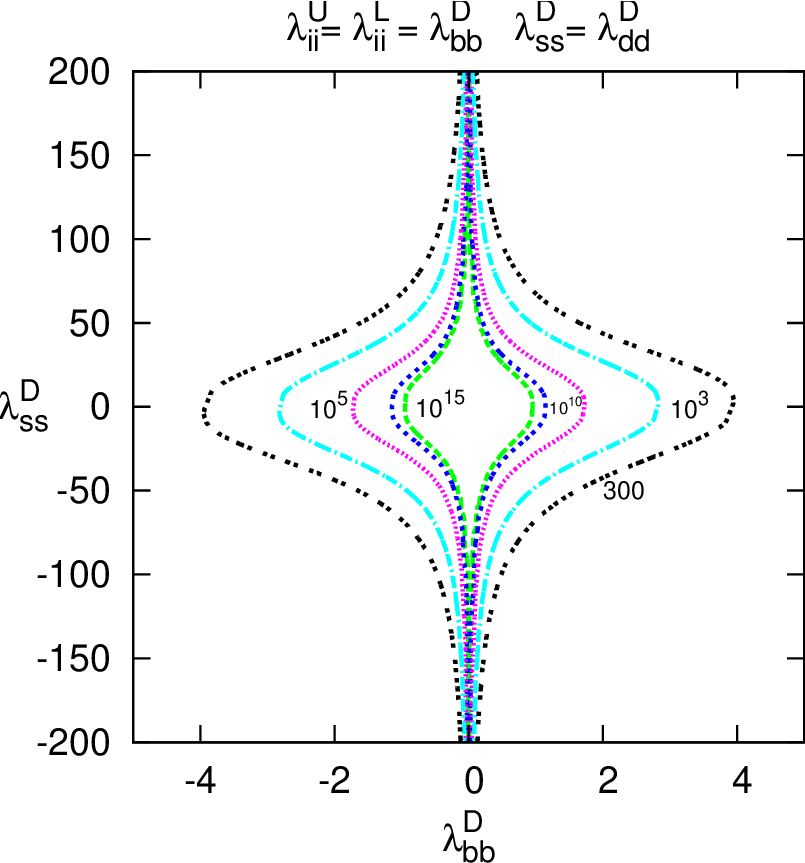} \hspace*{5mm}
\includegraphics[width=4.8cm]{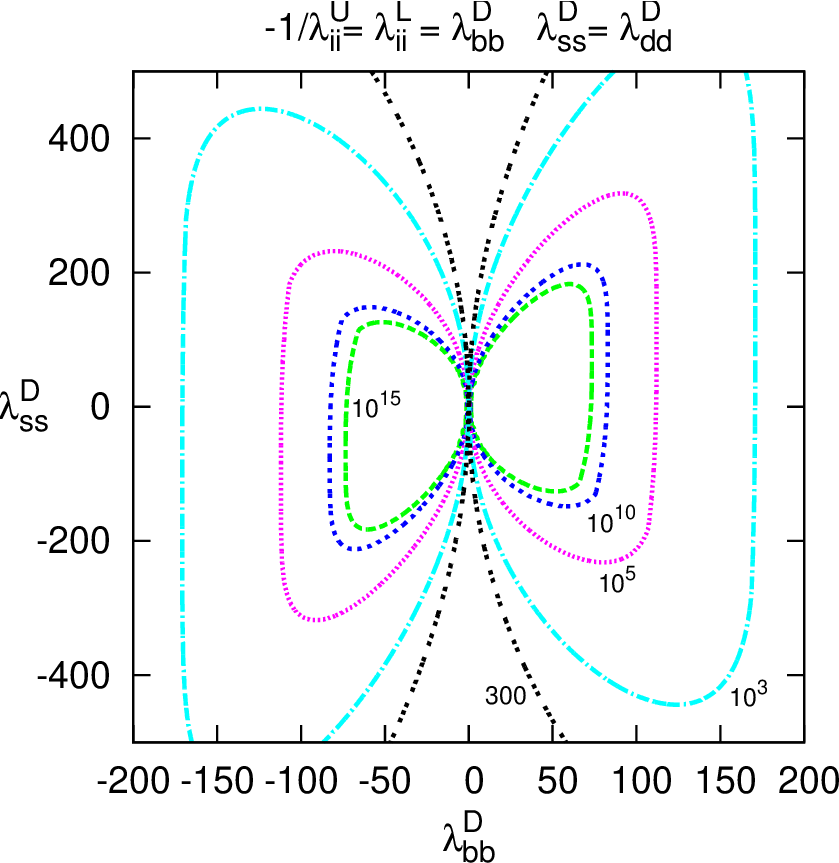}
 \caption{The energy scale where the Landau pole or non-diagonal $ \lambda^F_{i \neq j}  =  0.1$  is reached as a function of $\lambda_{ss}$ and $\lambda_{bb}$. In the left (right) panels $\lambda^{U}=\lambda_{bb} (-1/\lambda_{bb})$ and in all cases $\lambda^{L}=\lambda_{bb}$. }
    \label{plot:2D_MSD}
\label{plot:2D_MSD}
\end{figure}

\subsubsection{Non-diagonal models}
In the end we consider the models of  $Z_2$ symmetry breaking from having non-zero
non-diagonal elements in the up or down sectors. We set
$\lambda^U_{i \neq j}=0.1 $ or $\lambda^D_{i \neq j}=0.1$ at the EW scale and then evolve to higher energy. The analysis shows the constraints from off-diagonal elements are small compare to the previous cases.
\begin{figure}
\centering
\includegraphics[width=4.6cm]{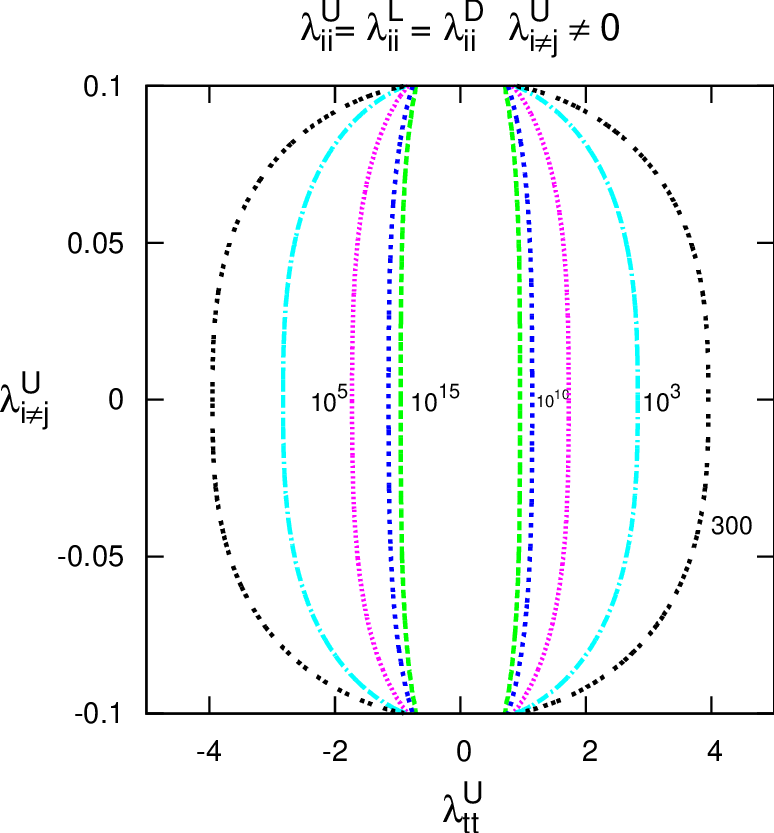}
\centering
\hspace{0.5cm}
\includegraphics[width=4.6cm]{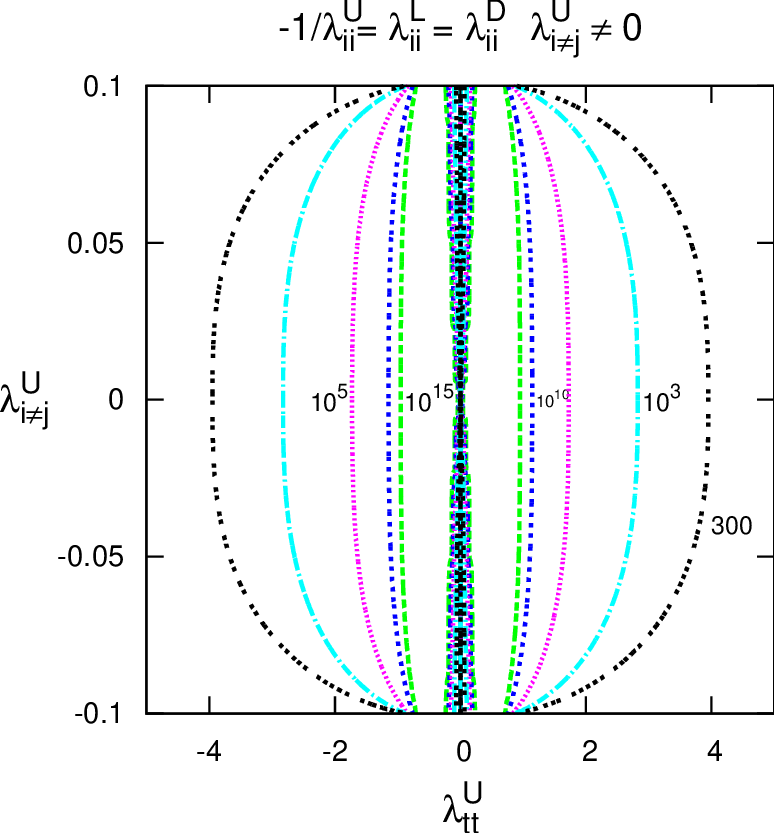} \newline \\
\hspace{-2.0cm}
\includegraphics[width=4.6cm]{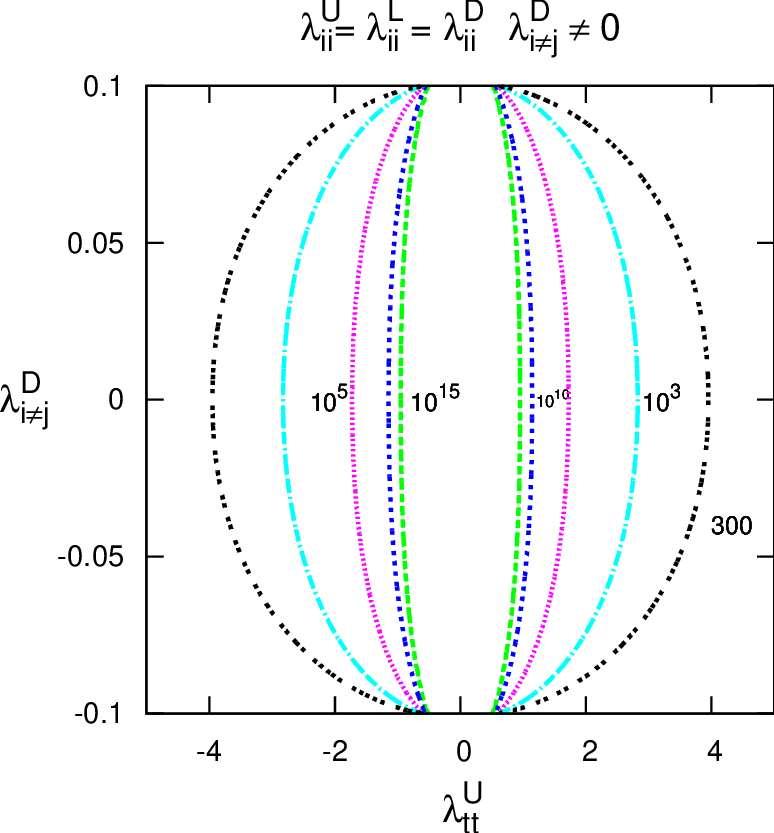}
\centering
\hspace{0.5cm}
\includegraphics[width=4.6cm]{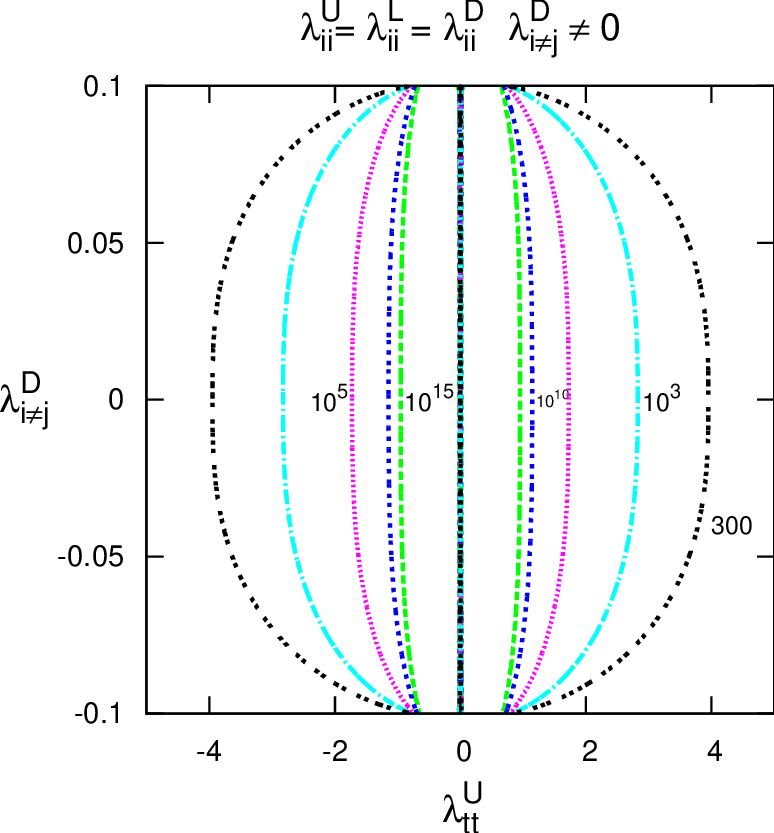}

  \caption{The constraints from the Landau pole and the off-diagonal elements as a function of $\lambda^U_{ii}$ and the off-diagonal elements $\lambda^U_{i \neq j}$ (up) or    $\lambda^D_{i \neq j}$ (down)  at the input scale for the type I (left) and type II (right) relations for the diagonal elements.}
\end{figure}

\section{Conclusions}
RGE evolution is a useful tool to analyse different 2HDM models on stability of underlying assumptions. A quick appearance of Landau pole or large off-diagonal Yukawa coupling under RGE evolution may indicate the
model is either fine tuned or incomplete, e.g., new particles appearing at high energy.


\begin{thebibliography}{99}

\bibitem{Glashow:1976nt}
  S.~L.~Glashow, S.~Weinberg,
  Phys.\ Rev.\  {\bf D15 } (1977)  1958.

\bibitem{Pich:2009sp}
 A.~Pich and P.~Tuzon,
 Phys.\ Rev.\  D {\bf 80} (2009) 091702
 [arXiv:0908.1554].

 \bibitem{Ferreira:2010xe}
  P.~M.~Ferreira, L.~Lavoura and J.~P.~Silva,
  Phys.\ Lett.\  B {\bf 688} (2010) 341
  [arXiv:1001.2561].

 \bibitem{Cheng:1987rs}
 T.~P.~Cheng and M.~Sher,
 Phys.\ Rev.\  D {\bf 35} (1987) 3484.

\bibitem{Atwood:1996vj}
  D.~Atwood, L.~Reina, A.~Soni,
  Phys.\ Rev.\  {\bf D55 } (1997)  3156-3176.
  [hep-ph/9609279].

\bibitem{Bijnens:2011gd}
  J.~Bijnens, J.~Lu and J.~Rathsman,
  JHEP {\bf 1205} (2012) 118
  [arXiv:1111.5760].

\end{thebibliography}
\end{document}